A damage-mechanics model for fracture nucleation and propagation

G. Yakovlev[1], J. D. Gran*[1], D.L. Turcotte[2], J.B. Rundle[123], J.R. Holliday[1], W. Klein[4]

*Corresponding Author
[1] Department of Physics, One Shields Ave, University of California, Davis, Davis CA, 95616
[2] Department of Geology, One Shields Ave, University of California, Davis, Davis CA, 95616
[3] Sante Fe Institute, Santa Fe, NM 87501
[4] Department of Physics, Boston University, Boston, MA 02215
glebos@gmail.com (G. Yakovlev)
gran@student.physics.ucdavis.edu (J.D. Gran)
dlturcotte@ucdavis.edu (D.L Turcotte)
rundle@physics.ucdavis.edu (J.B. Rundle)
holliday@physics.ucdavis.edu (J.R. Holliday)
klein@bu.edu (W. Klein)

Abstract
In this paper a composite model for earthquake rupture initiation and propagation is proposed. The model includes aspects of damage mechanics, fiber-bundle models, and slider-block models. An array of elements is introduced in analogy to the fibers of a fiber bundle. Time to failure for each element is specified from a Poisson distribution. The hazard rate is assumed to have a power-law dependence on stress. When an element fails it is removed, the stress on a failed element is redistributed uniformly to a specified number of neighboring elements in a given range of interaction. Damage is defined to be the fraction of elements that have failed. Time to failure and modes of rupture propagation are determined as a function of the hazard-rate exponent and the range of interaction.

1. Introduction
A composite model is introduced in this paper for the nucleation and propagation of fractures. The model incorporates aspects of damage mechanics, fiber-bundle models, and slider-block models. A square array of elements is considered, these elements are analogous to the fibers in a fiber-bundle model and the blocks in a slider-block model. At time $t=0$ a constant force is applied to the system. Time-to-failure statistics are prescribed. When an element fails the stress on that element is transferred to a prescribed range of adjacent elements. Numerical simulations are used to study the conditions under which a well defined rupture nucleates and to illustrate the propagation of this fracture over the array.
The model is closely related to the fiber-bundle model. The fiber bundle initially consists of $n_0$ fibers. Subsequently $n_f$ fibers fail and when $n_f=n_0$ the bundle fails. When a fiber fails the load on that fiber is transfered to other fibers. In the equal load sharing case the load is transfered to all other fibers equally. In the local load sharing case the load is transfered to the adjacent fibers within a prescribed interaction region. Two failure criteria have been proposed. The first is static and a failure strength is prescribed statistically for each fiber[1]. As the stress on the fibers increase, more fibers fail. The second failure criterion specifies a statistical time to failure for each fiber that is stress dependent[2,3]. In terms of applicability the latter approach is now generally accepted. A general review of fiber-bundle models has been given in[4].
Damage mechanics is an empirical continuum approach to material failure[5,6]. A continuum damage variable $α$ is defined by the relation

$$E = E_0(1-\alpha) \tag{1}$$

where $E$ is the Young's modulus for the damaged material and $E_0$ is the Young's modulus for the undamaged material. When $\alpha=1$ failure occurs. A rate equation for the increase in damage is specified. There is a close association between the equal-load sharing fiber-bundle model and damage mechanics if it is assumed that[7]

$$\alpha = \frac{n_f}{n_0} \tag{2}$$

Damage mechanics does not consider the propagation of a rupture.

Slider-block models have been studied extensively as models for earthquakes[8,9,10,11,12]. An array of slider blocks is pulled along a surface with puller springs. When the stress on a block exceeds the static coefficient of friction it slips and stress is transfered to other blocks by connector springs. Extensive studies of the role of stress transfer have been carried out[13]. When a block fails the stress on the block is redistributed equally to neighboring blocks in a given range of interaction. This approach has been used to study a slider-block model in a failure mode [14]. However, no time to failure statistics were incorporated.

2. The Model

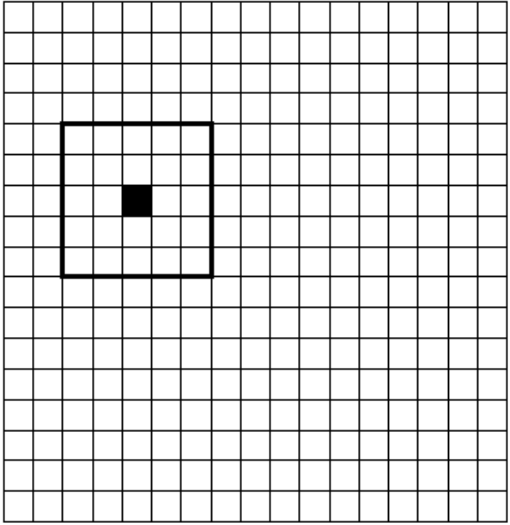

Figure 1: Illustration of the model. A square grid of elements with width $L$ is considered, in this case $L=17$ so that there are $n_0 = 17^2 = 289$ elements. At time $t=0$ a uniform load $F_0$ is uniformly distributed to all the elements. The stress $\sigma_0$ on each element is $\sigma_0 = F_0/n_0$. One element has the shortest time to failure $t_{f,min}$, the failure of this element is the solid square. The stress from this failed element is redistributed equally to all the elements within range $r$. Here $r=2$ and is represented by the black line surrounding the failed element. The stress is redistributed equally to the 24 remaining elements within the square.

An illustration of the model is given in Fig. 1. A square grid of elements is considered with L elements on a side, for the case illustrated $L=17$ so that the total number of elements is $n_0 = 17^2 = 289$. At time $t=0$ a constant load $F_0$ is applied to the grid and it is uniformly distributed so that each element has a stress

$$\sigma_0 = \frac{F_0}{n_0} \tag{3}$$

No additional external forces are applied to the system. A time to failure $t_f$ is assigned randomly to each element from a prescribed distribution. In this paper it is assumed that this distribution is Poisson so that there is no memory of the stress history of an element. The cumulative history of failure times is thus given by

$$P_c(t_f) = 1 - e^{-\nu t_f} \tag{4}$$

where $\nu$ is the hazard rate. This distribution has been shown to be applicable to the distribution of nucleation times in solid-state physics, specifically the Ising model[15]. A second assumption is that the hazard rate $\nu(\sigma)$ has a power-law dependence on the stress $\sigma$ on the element

$$\nu(\sigma) = \nu_0 \left(\frac{\sigma}{\sigma_0}\right)^\rho \tag{5}$$

where the power-law exponent $\rho$ must be specified and $\nu_0$ is the value of the hazard rate when $\sigma = \sigma_0$. It is found experimentally that values of $\rho$ are in the range 2-5 for various fibrous materials[16]. Initially, at $t=0$, the stresses on all elements are equal with the value $\sigma_0$ given in Eq.(3). For each element a random number $P_c$ in the range 0 to 1 is chosen. Using this random number the corresponding failure time of the element is obtained from Eq.(4) with $\nu = \nu_0$

$$t_f = \frac{1}{\nu_0} \ln[(1-P_c)^{-1}] \tag{6}$$

The first element fails at $t = t_{f,min}$ the smallest of these failure times. This first failed element is illustrated in Fig. 1. The failed element is removed from the grid (there is no healing). The stress on the failed element is redistributed equally to all surviving elements in a range of interaction $r$. For the example illustrated in Fig. 1 the range of interaction is $r=2$. The redistribution is carried out over a square region with $2r+1$ elements on a side. The maximum number of elements $n_{rd}$ over which the stress is redistributed is

$$n_{rd} = (2r+1)^2 - 1 \tag{7}$$

For the example in Fig. 1, $n_{rd}=24$. In subsequent redistributions some of the elements in the region may have been removed due to previous failures. In this case the stress is redistributed equally to the surviving elements. If the failed element has no surviving neighbors, the stress on that element is dissipated from the system, reducing the total load.

All surviving elements in the grid are given a new time to failure $\Delta t_f$ that is determined from Eqs.(4) and (5) written in the form

$$\Delta t_f = \frac{1}{\nu_0}\left(\frac{\sigma_0}{\sigma}\right)^\rho \ln[(1-P_c)^{-1}] \tag{8}$$

where $P_c$ is a new random number in the range 0 to 1. This approach is appropriate for the Poisson distribution of failures given by Eq. (4) since a surviving element has no memory of the prior stress history. Considering the values $\Delta t_f$ for all elements, the shortest time to failure is determined. At this time this failed element is removed from the grid. This process is continued until all elements have failed. This is the failure time $t_{gf}$ for the grid. At this time the number of failed elements $n_f$ is equal to the number of elements originally on the grid $n_0$, $n_f = n_0$. Following the standard association of damage mechanics with the fiber-bundle model we take the damage variable $\alpha$ to be given by Eq.(2). The damage variable is the fraction of failed elements, failure of the grid occurs at $\alpha = 1$. A primary object of our simulations is to determine $\alpha$ as a function of $t$ ($0 \leq t \leq t_{gf}$).

3. Mean Field Analysis

The case in which the stress on a failed element is redistributed equally to the surviving elements on the grid can be solved analytically. This is known as equal-load sharing and is the mean-field limit for this problem. For this case the stresses on all surviving elements $\sigma_{mf}(t)$ are equal. The condition that the total force $F_0$ on the grid remains constant for $t>0$ requires

$$(n_0 - n_f)\sigma_{mf} = n_0 \sigma_0 \tag{9}$$

The standard breakdown rule for the rate of failure of fibers given in [2,3,4] is

$$\frac{d(n_0 - n_f)}{dt} = -v(\sigma)(n_0 - n_f) \tag{10}$$

where $v$ is again the hazard rate. If $v$ is a constant $v_0$ the integration of Eq. (10) gives the probability distribution given in Eq. (4). It is important to note that there is a correspondence between the fiber-bundle formulation and the damage formulation only if the Poissonian failure condition in Eq. (4) is used.

The power-law dependence of the hazard rate on stress as given in Eq.(5) is assumed to be valid. Combining Eqs. (5) and (9) gives

$$v(n_f) = v_0 \left(1 - \frac{n_f}{n_0}\right)^{-\rho} \tag{11}$$

and substitution into Eq. (10) gives

$$\frac{d(n_0 - n_f)}{dt} = \frac{-v_0 n_0^\rho}{(n_0 - n_f)^{\rho - 1}} \tag{12}$$

Integrating with $n_f=0$ at $t=0$ with the definition of the damage variable $\alpha$ given in Eq. (2) gives

$$\alpha = \frac{n_f}{n_0} = 1 - (1 - v_0 \rho t)^{\frac{1}{\rho}} \tag{13}$$

And grid failure occurs at time

$$t_{gf} = \frac{1}{v_0 \rho} \tag{14}$$

It should be noted that the results given above are applicable for large values of $n_0$ [4]. Numerical simulations are well approximated by the above results in the equal load sharing (mean-field) limit. These results demonstrate that the appropriate non-dimensional time for our model is $v_0 t$.

In the equal load sharing (mean-field) limit failures are spatially random. There is no spatial localization of failure and thus no initiation and propagation of a fracture. In this limit the model is identical to the site-percolation model[17]. This model also consists of a square grid of elements. Percolating elements $n_p$ are picked at random with the probability $p=n_p/n_0$ a variable. This is also a transient problem with increasing $p$. When $p=p_c$ when a percolating cluster of elements spans the grid[18]. This is a well documented critical point. However, this critical point is not relevant to the total grid failure at $n_p=n_0$.

4.   Simulations

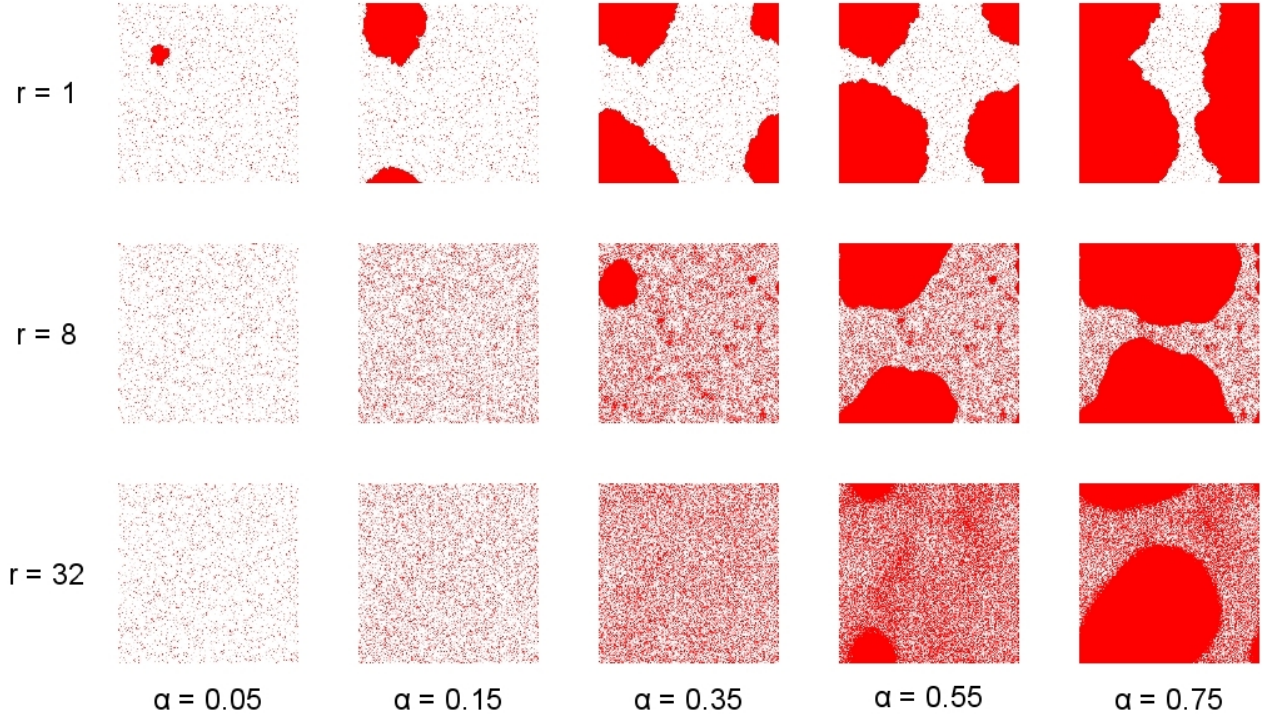

Figure 2: Three examples of rupture nucleation and propagation are illustrated. We assume $\rho = 3$, $L=512$, and consider $r = 1$, $8$ and $32$. For each case we give the structure for $\alpha=0.05, 0.15, 0.35, 0.55$ and $0.75$. The dark areas are failed elements.

In order to illustrate the behavior of this model a sequence of numerical simulations have been carried out. In order to carry out a simulation it is necessary to specify: (1) The size of the grid $n_0$, (2) the range of interaction $r$, and (3) the power-law exponent $\rho$.

In Fig. 2 the nucleation and propagation of fractures is illustrated for several typical examples. The power-law exponent introduced in Eq. (5) is assumed to have the value $\rho=3$. In addition, the array has a width $L=512$ elements so that the total number of elements is $n_0=512^2=262,144$. Three examples are given with the ranges of interaction $r=1, 8$ and $32$. The distributions of failed sites (damage) are illustrated for values of the damage variable (fraction of failed elements) $\alpha=0.05, 0.15, 0.35, 0.55$ and

0.75. For the nearest neighbor case, *r=1*, the fracture has just started to nucleate at *α=0.05*. It subsequently spreads across the array. Note periodic boundary conditions are assumed so that when the propagating fracture reaches the top boundary of the grid its extension appears at the lower boundary. The propagating fracture is quite irregular and has many indentations. For the intermediate range case *r=8* the incipient rupture is delayed until *α=0.35*. Again the spread of this rupture across the array is clearly illustrated. The rupture in this case is more nearly circular with smoother boundaries. There is also a clearly defined halo of failed elements adjacent to the boundary associated with the *r=8* range of stress transfer. For the near mean-field case with *r=32* rupture initiation is delayed further and occurs at about *α=0.55*. The rupture has a diffuse outer boundary as it propagates across the remainder of the array.

In the three examples a nucleation phase is followed by a propagation phase. In the mean-field limit *r=64* there is no propagation phase. Failed elements are randomly distributed across the grid.

(a)
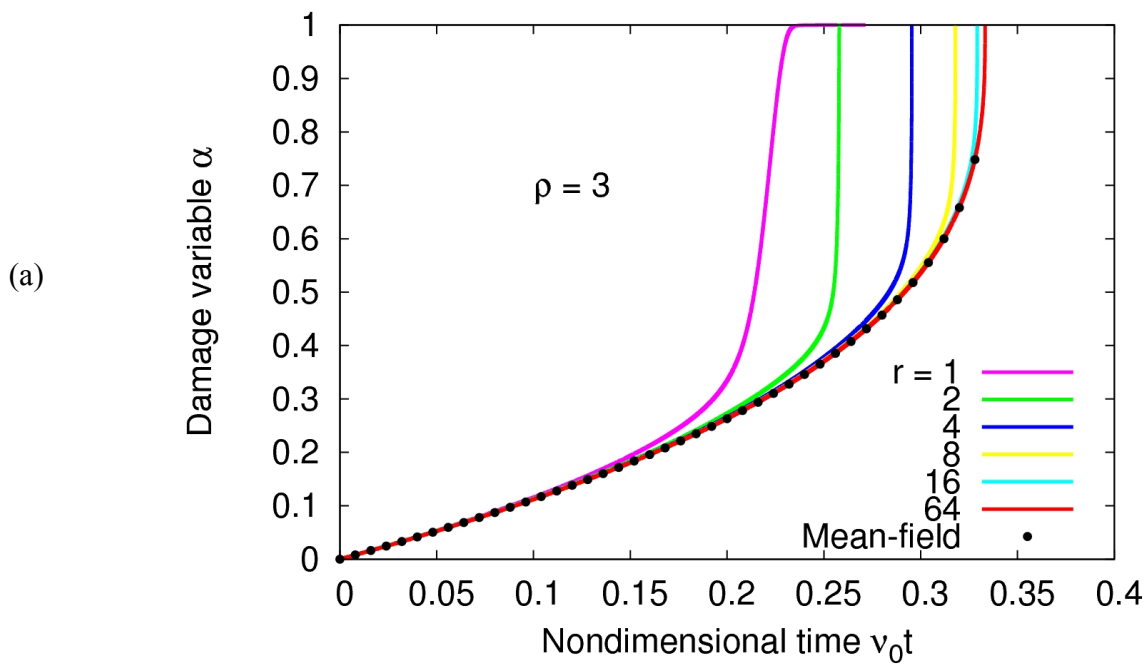

(b)
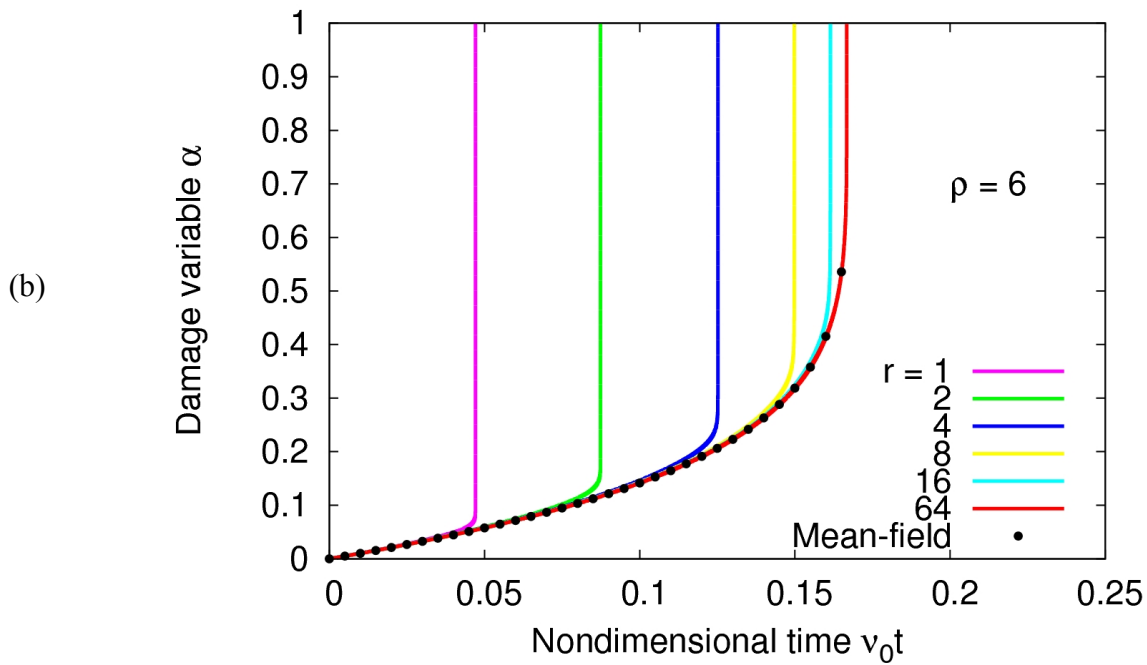

Figure 3: Dependence of the damage variable $\alpha$ on the non-dimensional time $v_0 t$ after the application of the constant applied force for (a) stress exponent $\rho=3$ and (b) stress exponent $\rho=6$. Results are given for ranges of interaction $r = 1,2,4,8,16$ and $64$ (mean field). Results are the average of 500 simulations. Also shown is the mean-field solution given in Eq. (13).

It is also of interest to consider the temporal evolution of the damage variable during rupture nucleation and propagation. Several examples are given in Fig. 3. It is important tot note that there is considerable variability in each individual simulations. In each case we carry out 500 simulations and the mean times to reach a specified value of the damage variable are obtained. The values of the damage variable $\alpha$ are given as a function of these mean non-dimensional times $v_0 t$ after the application of the constant force are given with $L=129$ and $n_0=16,641$ and $\rho=3$ in Fig. 3(a) and $\rho=6$ in Fig. 3(b). Results are given for a sequence of values for the range of interaction from $r=1$ (nearest neighbor) to $r=64$ (mean-field). For small values of r the transition from the nucleation phase to the propagation phase is clearly illustrated . For the mean field limit $r=64$ the evolution of the damage variable is in agreement with Eq.(13). For $\rho=3$ the failure takes place at $v_0 t_{gf}=1/3$ and for $\rho=6$ at $v_0 t_{gf}=1/6$ as predicted by Eq. (14). The behavior for the two values of $\rho$ is quite similar except for the reduction in the time to failure for larger $\rho$ as well as a sharper transition from nucleation to propagation.

For all values of the interaction range r for a fixed $\rho$ the initial increase of damage is essentially identical. The transition from rupture nucleation to propagation is governed by the range of interaction. The transition time increases systematically with increasing range of interaction. For a short range of interaction the stress level adjacent to a rupture is greater, increasing the rupture speed and subsequently decreasing the time to failure. When the rupture speed approaches the velocity of shear waves, dynamic affects become important which are not included in this analysis. These effects, however, will not change the times to failure which are dominated by the nucleation phase.
The transition from the nucleation phase to the propagation phase is quantified using the maximum curvature of the curves for $\alpha(t)$ plotted in Fig. 3. The curvature $k$ is given by

$$k = \frac{\frac{d^2 \alpha}{dt^2}}{[1+(\frac{d\alpha}{dt})^2]^{\frac{3}{2}}} \tag{15}$$

The non-dimensional transition times $v_0 t_t$ corresponding to the maximum values of $k$ are given in Table 1. For $\rho=3$ the values decrease systematically for increasing $r$. For $\rho=6$ the maximum value is at $r=2$.

| $r$ | $\rho=3$ | $\rho=6$ |
|---|---|---|
| 1 | 0.204 | 0.047 |
| 2 | 0.252 | 0.087 |
| 4 | 0.292 | 0.125 |
| 8 | 0.313 | 0.149 |
| 16 | 0.327 | 0.157 |
| 32 | 0.330 | 0.159 |

Table 1: Non-dimensional times of transition from the nucleation phase to propagation phase are given for several values of the interaction range $r$ and stress exponent $\rho$. The transition times are computed by locating the point of maximum curvature in the curves plotted in Fig. 3(a) and 3(b).

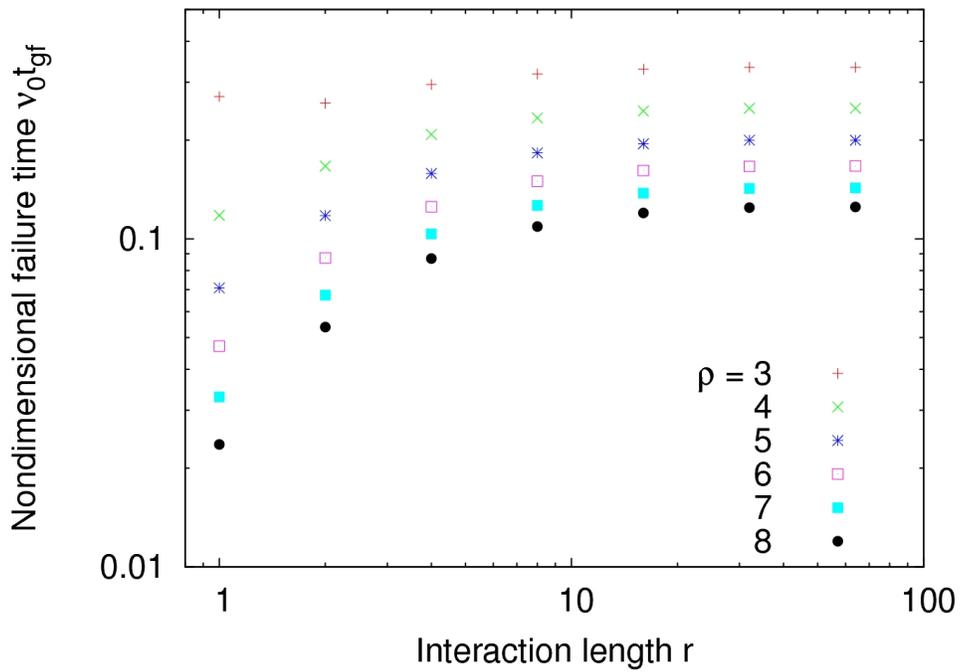

Figure 4: Dependence of the non-dimensional time to failure $v_0 t_{gf}$ on the interaction length $r$. Results are given for stress exponents $\rho=1,2,3...8$ and are the average of 500 simulations.

The non-dimensional times to failure $v_0 t_{gf}$ is given as a function of the range of interaction $r$ ($r$ = 1,2,4,8,16,32 and 64,mean field) in Fig. 4. Results are given for the power-law exponent values $\rho$ = 3,4,5...8. The failure times decrease with increasing values of $\rho$ as would be expected. As the range of interaction is increased for fixed $\rho$ the failure times asymptotically approach the mean-field limit. For $r=64$ (mean field) the results are in agreement with the mean-field theory given in Eq. (14). Again, each value is the average of 500 simulations.

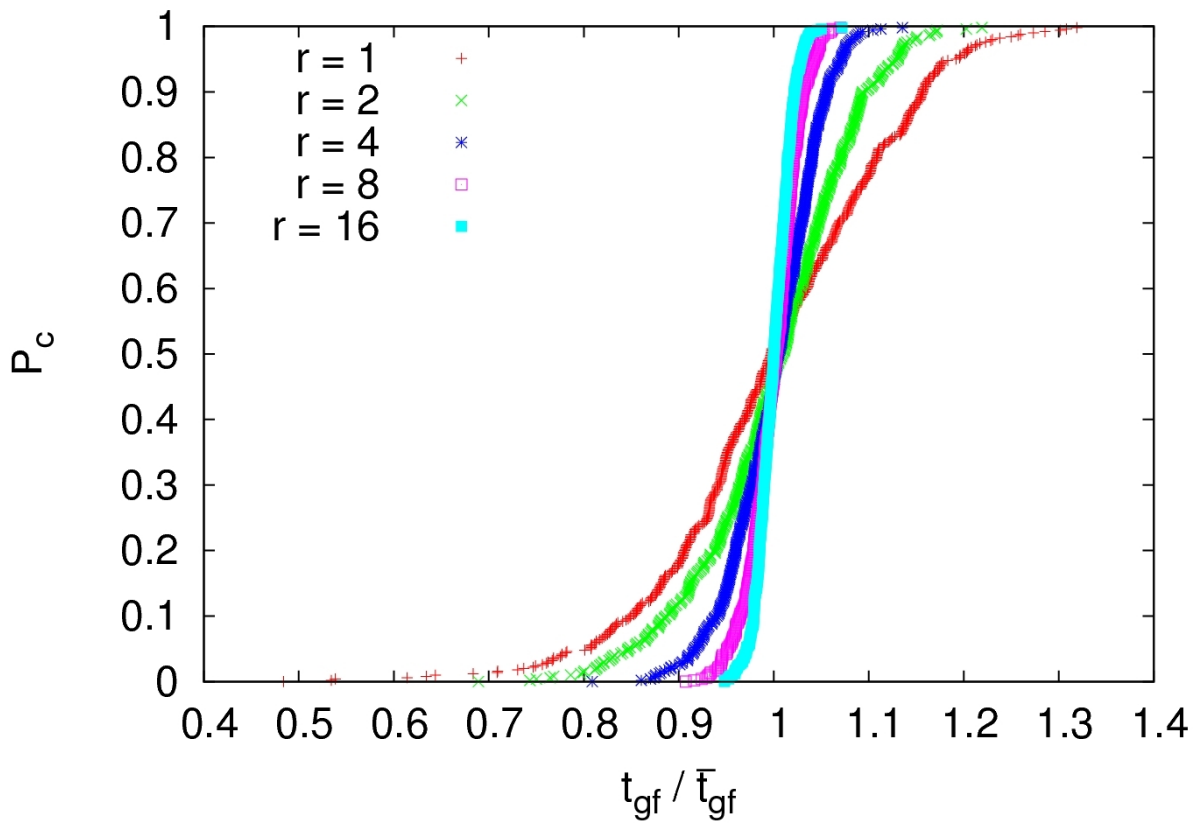

Figure 5: Cumulative distribution function $P_c$ of failure times $t_{gf}/\bar{t}_{gf}$ for several values of the interaction length $r$, for $\rho=6$.

In Fig. 5 the statistical distributions of failure times are given. The cumulative distribution function of failure times $P_c$ is given as a function of the ratio of the non-dimensional failure time to its mean value $t_{gf} = \bar{t}_{gf}$. This dependence is given in Fig. 5 for $\rho=6$ and $r = 1,2,4,8$ and $16$. In the mean field limit ($r = 64$) the standard deviation of the failure times goes to zero. The standard deviation increases systematically with decreasing values of $r$.

It is also of interest to study the dependence of the damage parameter $\alpha$ on the time prior to failure $v_0(t_{gf} - t)$. This is given in Fig. 6 for $\rho=6$ and several values of $r$. For $r = 64$ the mean field result given in Eq. (13) is found with the power law dependence

$$1-\alpha = \left(1 - \frac{t}{t_{gf}}\right)^{\frac{1}{6}} \tag{3}$$

As r approaches the mean field limit the initial power-law behavior of becomes apparent. For $r=1$ the initial power-law behavior is absent. Again, these results are averages over 500 simulations.

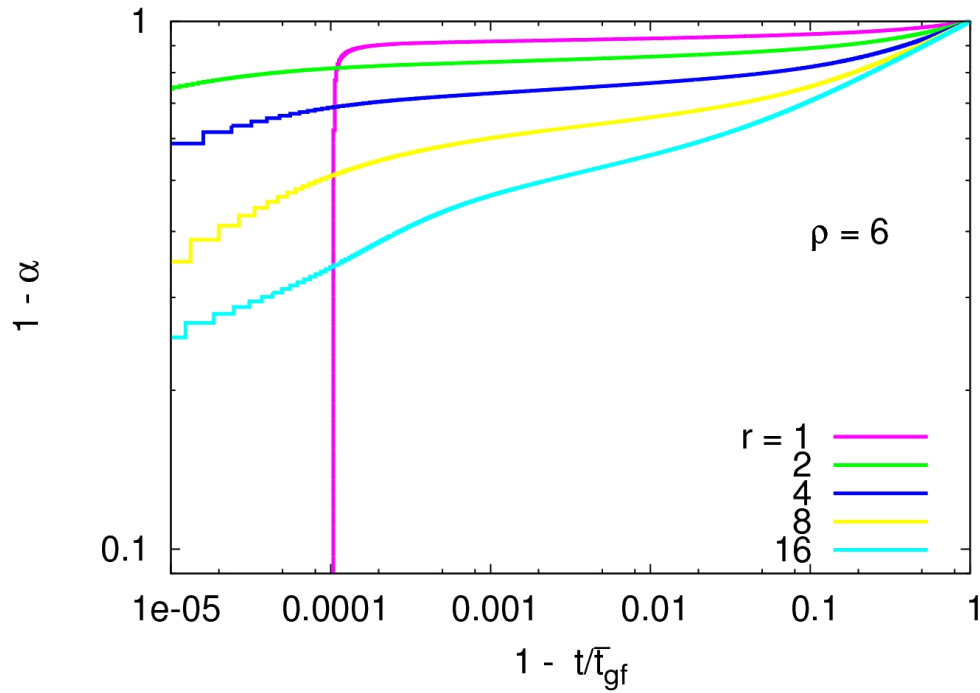

Figure 6: Dependence of $1 - \alpha$ on the time prior to failure $1 - t/\bar{t}_{gf}$ for $\rho=6$ and several values of $r$.

5. Discussion

The simulations given above show a wide range of behaviors from gradual failure to catastrophic failure. The principle control parameter is the range of interaction $r$. When the range of interaction is equal to the system size the model is in equal load sharing mode. When an element fails the stress on the element is transferred equally to all surviving elements. This is the mean-field limit for this problem. The analytic solution for this problem has been given in Eqs. (13) and (14). A gradual power-law increase in damage is found. There is no fracture propagation. All systems fail at about the same time, the standard deviation of failure times is small. The distribution of failed elements is identical to the distribution of percolating sites in site percolation. The other extreme of the behavior is nearest neighbor load sharing with $r=1$. For the case illustrated in Fig. 2, the failure patch is well defined with the $\alpha=0.05$ value of the damage parameter. Over 95% of the damage is associated with the propagating fracture. The time to failure is also much more rapid than compared to the mean-field case. From Fig. 4 and with $\rho=6$ the time to failure is $v_0 t_{gf}=0.047$, this compares with an equal load sharing value $v_0 t_f = 1/3$. There is also a relatively large range of failure times as shown in Fig. 5.

Acknowledgments
Research by G.Y., J.D.G., J.B.R. J.R.H., and D.L.T. has been supported by a grant from the U.S. Department of Energy, Office of Basic Energy Sciences to the University of California, Davis, DE-FG02-04ER15568. The research by W.K. has been supported by a grant from the U.S. Department of Energy, Office of Basic Energy Sciences to Boston University, DE-FG02-95ER14498.

References

[1] H. E. Daniels. The statistical theory of the strength of bundles of threads. I. *Proc. Roy. Soc. Lond. A,* 183:405-435, 1945. doi: 10.1098/rspa.1945.0011.
[2] B. D. Coleman. Time dependence of mechanical breakdown in bundles of fibers. I. Constant total load. *J. Ap. Phys.*, 28:1058-1064, 1957.
[3] B. D. Coleman. Statistical and time dependence of mechanical breakdown in fibers. *J. Ap. Phys.*, 29:968-983, 1958.
[4] W. I. Newman and S. L. Phoenix. Time-dependent fiber bundles with local load sharing. *Phys. Rev. E*, 63:021507, 2001.
[5] D. Krajcinovic. *Damage Mechanics*. Elsevier, Amsterdam, 1996.
[6] R. Shcherbakov and D. L. Turcotte. Damage and self similarity in fracture. *Theor. Ap. Frac. Mech.*, 39:245-258, 2003.
[7] D. L. Turcotte, W. I. Newman, and R. Shcherbakov. Micro and macroscopic models of rock fracture. *Geophys. J. Int.*, 152:718-728, 2003.
[8] R. Burridge and L. Knopoff. Model and theoretical seismicity. *Bull. Seismol. Soc. Am.*, 57:341-371, 1997.
[9] J. B. Rundle and D. D. Jackson. Numerical simulation of earthquake sequences. *Bull. Seismol. Soc. Am.*, 67:1363-1377, 1977.
[10] J. M. Carlson and J. S. Langer. Mechanical model of an earthquake fault. *Phys. Rev. A*, 40:6470-6484, 1989.
[11] J. B. Rundle and W. Klein. Scaling and critical phenomena in a cellular automaton model for earthquakes. *j. Stat. Phys.*, 72:405-412, 1993.
[12] J. B. Rundle, D. L. Turcotte, R. Shcherbakov, W. Klein, and C. Sammis. Statistical physics approach to understanding the multiscale dynamics of earthquake fault systems. *Rev. Geophys.*, 41:1019, 2003
[13] W. Klein, M. Anghel, C. D. Ferguson, J. B. Rundle, and J. S. Sa Martins. Statistical analysis of a model for earthquake faults with long-range stress transfer. In J.B. Rundle, D.L. Turcotte, and W. Klein, editors, *GeoComplexity and the Physics of Earthquakes*, Geophysical Monograph 120, pages 43-71. American Geophysical Union, Washington D.C., 2000.
[14] C. A. Serino, W. Klein, and J. B. Rundle. Cellular automaton model of damage. *Phys. Rev. E*, 81(1):016105, Jan 2010.

[15] K. Brendel, G. T. Barkema, and H. van Beijeren. Nucleation times in the two dimensional ising model. *Phys. Rev. E,* 71:031601, 2005.

[16] W. A. Curtin and H. Scher. Time-dependent damage evolution and failure in materials. I. Theory. *Phys. Rev. B*, 55:12038-12050, 1997.

[17] D. Stauffer and A. Aharony. *Introduction to Percolation Theory.* Taylor and Francis, London, second edition, 1992.

[18] D. L. Turcotte. Self-organized criticality. *Rep. Prog. Phys.* 62:1377-1429, 1999.